\documentclass[onecolumn,aps,floatfix,pra,superscriptaddress,nofootinbib]{revtex4}
\usepackage{hyperref}
\usepackage{color}
\usepackage{graphicx}
\usepackage{bm}
\usepackage{amsmath}
\usepackage{hyperref}
\usepackage{enumerate}
\usepackage{upgreek}
\usepackage[subnum]{cases}

\newcommand{\ep}{ \epsilon }
\newcommand{\pa}{ \partial }
\newcommand{\hb}{ \hbar }
\newcommand{\si}{ \sigma }

\newcommand{\ga}{ \gamma }

\newcommand{\hbt}{ \widetilde{\hbar} }
\newcommand{\rhot}{ \widetilde{\rho} }
\newcommand{\psit}{ \widetilde{\psi} }
\newcommand{\wt}{ \widetilde{w} }
\newcommand{\Nt}{ \widetilde{\mathcal{N}} }
\newcommand{\Gat}{ \widetilde{\Gamma} }
\newcommand{\Pt}{ \widetilde{P} }
\newcommand{\Qefft}{ \widetilde{\mathcal{Q}} }

\newcommand{\jt}{ \tilde{j} }
\newcommand{\bzet}{ \tilde{b}_0 }
\newcommand{\bont}{ \tilde{b}_1 }

\newcommand{\res}{ \mathrm{res} }

\begin{document}
\title{ Scaled Caldeira-Leggett Dynamics: Deterministic Trajectories and the Classical Limit }
\author{S. V. Mousavi}
\email{vmousavi@qom.ac.ir}
\affiliation{Department of Physics, University of Qom, Ghadir Blvd., Qom 371614-6611, Iran}
\author{S. Miret-Art\'es}
\email{s.miret@iff.csic.es}
\affiliation{Instituto de F\'isica Fundamental, Consejo Superior de Investigaciones Cient\'ificas, Serrano 123, 28006 Madrid, Spain}

\begin{abstract}

The dynamics of an open quantum system in the high-temperature regime of the Caldeira-Leggett model using a Bohmian language is investigated. By expressing the density matrix in polar form,  generalized continuity and Hamilton-Jacobi equations are obtained. Thermal effects appear explicitly in the former, while they influence the latter only indirectly through an effective potential. Remarkably, the effective potential does not vanish in the classical limit and cannot, in general, be decomposed into simple additive quantum and thermal contributions. Instead, it retains a nontrivial structure leading to a residual temperature dependence in the resulting dynamics.
As a consequence, the resulting temperature-dependent-classical trajectories remain deterministic even in the presence of a thermal environment. This behavior contrasts with the stochastic dynamics observed in the Langevin description and reflects the ensemble-based nature of the present approach. To further explore the quantum-to-classical transition, a scaled version of the Caldeira-Leggett equation is introduced by scaling both the density matrix and Planck constant through a quantumness parameter. This parameter takes the value one in the fully quantum regime and smoothly approaches zero in the classical limit. Applications to Gaussian states demonstrate a gradual suppression of coherence and interference, governed jointly by thermal effects and the transition parameter.

\end{abstract}

\maketitle



\section{Introduction}

The classical limit of quantum mechanics remains an open problem, even after nearly a century since the formulation of the theory. The limit of a vanishing Planck constant is known to be both mathematically subtle and conceptually unclear \cite{Holland-book-1993, Ro-FP-1986}. By expressing the wave function in polar form within the Schrödinger equation and separating the resulting equation into its real and imaginary parts, one obtains, respectively, the Hamilton--Jacobi (HJ) equation supplemented by an additional term, the so-called quantum potential, and the continuity equation \cite{Bo-PR-1952}. Motivated by this procedure, Rosen \cite{Ro-AJP-1964} proposed a nonlinear classical Schrödinger equation in which the quantum potential is subtracted from the external potential. In this way, the classical HJ equation is recovered.

In an attempt to describe both quantum and classical mechanics within a unified framework, a nonlinear transition equation has been introduced. This equation incorporates a quantumness parameter that takes the value one in the fully quantum regime and vanishes in the classical limit. It has been shown that this nonlinear equation is equivalent to a linear, scaled formulation involving a scaled Planck constant and a scaled wave function \cite{Rietal-PRA-2014}. In this theoretical analysis, quantum-like features persist even in the transition regime, providing a smooth transition between quantum and classical behavior.
Since then, a number of works have explored different aspects of this approach. For instance, scaled trajectories have been used to investigate interference phenomena, showing that signatures of quantum behavior can still persist beyond the strictly quantum domain \cite{Ch-AP-2016}, as well as to study wave packet scattering from an Eckart barrier \cite{Ch-AP-2018}. 
The linear scaled Schrödinger equation has also been employed to analyze several fundamental problems. These include tunneling through a rectangular potential, total reflection from a hard wall, and the gravitational weak equivalence principle \cite{Mo-IJPR-2020}. In addition, diffraction in time and early arrival effects have been investigated in \cite{Mo-IJPR-2021}. 

The quantum-classical transition (QCT) in dissipative systems has also been considered within a variety of frameworks. In the Caldirola-Kanai model, studies have focused on wave packet propagations, arrival-time distributions, and actual momentum distributions \cite{MoMi-JPC-2018}, together with the classical limit of dissipative backflow \cite{MoMi-EPJP-2020_2}. Closely related problems have been addressed within the Schrödinger-Langevin equation, where phenomena such as tunneling \cite{MoMi-AP-2018} and wave packet scattering from a barrier \cite{Ch-IJQC-2019} have also been analyzed. For a review see, for example, Ref.  \cite{MoMi-FP-2022}.
In addition, the QCT has been extended to mixed states within the framework of the scaled von Neumann equation \cite{MoMi-Sy-2023} guided by the standard procedure \cite{Liboff2003}. Within the same setting, the bouncing ball problem, in both gravitational and harmonic potentials, has also been investigated \cite{MoMi-EPJP-2024}. More recently, by allowing the quantumness parameter to take values larger than one, so-called superquantum effects have been explored from a hydrodynamic perspective \cite{Ch-AP-2024}.

Understanding how classical behavior emerges from quantum dynamics in the presence of an environment still remains unsolved
in the theory of open quantum systems. The Caldeira--Leggett (CL) model \cite{CaLe-PA-1983, Caldeira-book-2014} provides a paradigmatic setting for addressing this question, as it captures the combined effects of dissipation and thermal fluctuations in a tractable form. 
The CL formalism has found numerous applications and extensions. Besides its original role in describing quantum Brownian motion \cite{LaMaLe-book-2019}, it has been formulated in phase space through a continued-fraction solution of the CL master equation and applied to Brownian motion in periodic potentials \cite{GaZu-JPA-2004}, shown to emerge naturally from the Bose--Fermi Kondo model with an Ohmic bath \cite{LiLeHo-PRL-2005}, reformulated as the adapted Caldeira--Leggett (ACL) model for finite-dimensional unitary simulations of decoherence \cite{AlBaaAr-PRR-2023}, and subsequently employed to investigate equilibration and thermalization \cite{Al-Entropy-2022} as well as the non-Markovian features \cite{MaSmVa-IJQI-2026}. It has further been extended to generalized Langevin dynamics under shear flow \cite{PeZa-PRE-2023}, non-equilibrium quantum thermodynamics with Gaussian reservoirs \cite{CaEs-Quantum-2026}, efficient real-time simulations based on the inchworm algorithm combined with the frozen Gaussian approximation \cite{WaYaCa-Quantum-2025}, driven CL baths and their dynamical response \cite{GrTh-PRE-2018}, investigations of positivity violations in the CL master equation \cite{HoBeLi-EPJD-2019}, the non-Markovian origin of dissipation \cite{Fe-PRA-2017}, vibrational spectroscopy with anharmonic system potentials \cite{GoIvKu-JCP-2016}, quantum dissipative dynamics in resonant tunneling diodes \cite{SaTa-NJP-2014}, recurrence phenomena and Loschmidt echoes \cite{BeKo-PLA-2013}, and the limitations of the CL model in describing phase transitions in superconducting circuits \cite{Kari-PRB-2024}.

Diffraction and interference of identical particles \cite{MoMi-EPJP-2020}, momentum-space decoherence of distinguishable and identical particles \cite{KhMoMi-En-2021}, the propagation of stretching Gaussian wave packets, arrival-time distributions, and diffraction in time \cite{MoMi-EPJP-2022}, modular variables and the limits of phase detectability \cite{Mo-PS-2026}, the dynamics of quantum correlations within the double CL formalism \cite{Mo-EPJP-2025}, and trajectory-based measures of nonlocality in the same framework \cite{Mo-PS-2025} constitute further representative applications of the CL formalism. Moreover, the quantum-to-classical transition in the CL framework has been investigated using the functional renormalization-group approach \cite{KoFaNaSa-AP-2017} and, more recently, has been explored in the context of surface diffusion \cite{Miret2026}.

In this work, we revisit QCT within CL framework by adopting a scaling strategy motivated by the dynamics of the full system-plus-environment. In the standard formulation of open quantum systems, the combined system of interest and its environment is described as an isolated entity obeying the von Neumann equation, while the reduced dynamics is obtained by tracing out the environmental degrees of freedom. Within the CL approach, where the environment is modeled as a harmonic bath, the use of the Born and Markov approximations together with an Ohmic spectral density leads, in the high-temperature limit, to the conventional CL equation.
In parallel with this structure, and motivated by recently proposed scaled von Neumann dynamics for isolated systems, we anticipate a corresponding scaled version of the CL equation by introducing a scaled density matrix and a scaled Planck constant. This construction does not follow from an explicit derivation within the present framework, but rather provides a consistent and physically motivated extension of the CL description to the QCT regime. The resulting formulation offers a unified setting in which dissipative dynamics and quantum-to-classical behavior can be analyzed in terms of observables across different regimes.

Our analysis focuses on both dynamical and coherence properties. We first investigate the structure of trajectories, highlighting how environmental effects enter differently in classical and non-classical regimes. We then quantify coherence loss using the $\ell_1$-norm for a single Gaussian wave packet and extend the study to a superposition of counter-propagating packets (cat states), where decoherence is directly visible through the suppression of interference fringes. This combined approach allows us to present a unified and physically transparent picture of the emergence of classicality in open quantum systems.

\section{Caldeira-Leggett equation: a Bohmian perspective and the scaled formalism}

For a particle of mass $m$ under the presence of an external potential $V(x)$, the well-known CL equation 
for the reduced density matrix in the position representation  reads \cite{CaLe-PA-1983}
\begin{eqnarray} \label{eq: CL eq}
\frac{\pa \rho(x, x', t)}{\pa t} &=& \left[ - \frac{\hb}{2mi} \left( \frac{\pa^2}{\pa x^2} - \frac{\pa^2}{\pa x'^2} \right) - \ga (x-x') \left( \frac{\pa}{\pa x} - \frac{\pa}{\pa x'} \right)
+ \frac{ V(x) - V(x') }{ i\hb } - \frac{D}{\hb^2} (x-x')^2 \right] \rho(x, x', t),
\end{eqnarray}
in the high-temperature limit, $D=2m\ga k_B T$ being the diffusion coefficient  and $k_B$, $\ga$ and $T$  the Boltzmann 
constant,  the friction coefficient and the temperature, respectively. In the usual center of mass and relative coordinates
\begin{numcases}~
R = \frac{x+x'}{2} , \label{eq: cm} \\
r = x - x'  ,  \label{eq: rel} 
\end{numcases}
Eq. (\ref{eq: CL eq}) can be rewritten as
\begin{eqnarray} \label{eq: CLeq_Rr}
\left[ \frac{\pa}{\pa t} - i \frac{\hb}{m} \frac{\pa^2}{\pa R \pa r} + \frac{ V(R+r/2) - V(R-r/2) }{ i\hb } + 2 \ga r \frac{\pa}{\pa r} + \frac{D}{\hb^2} r^2 \right] \rho(r, R, t) &=& 0 .
\end{eqnarray}

In the following subsections, we are going to analyze the CL differential equation in terms of a Bohmian language and in the so-called scaled theory \cite{MoMi-Sy-2023}.

\subsection{A Bohmian perspective}

It is very illustrative to consider the polar form of the reduced density matrix expressed as
\begin{eqnarray} \label{eq: rho-polar}
\rho(R, r, t) &=& A(R, r, t) e^{i S(R, r, t) / \hb},
\end{eqnarray}
and  to insert it into  Eq. \eqref{eq: CLeq_Rr} and then separate the resulting expression into its imaginary and real parts 
leading to a differential equation for the phase, $S(R, r, t)$,
\begin{eqnarray} \label{eq: HJ}
- \frac{\pa S}{\pa t} &=& \frac{1}{m} S_r S_R + 2 \ga r S_r + \mathcal{V+Q}, 
\end{eqnarray}
and for the amplitude, $A(R, r, t)$,
\begin{eqnarray} \label{eq: con}
\frac{\pa A}{\pa t} + \frac{\pa}{\pa R} \left( A \frac{1}{m} S_r \right) + \frac{1}{m} A_r S_R + 2 \ga r A_r + \frac{D}{\hb^2} r^2 A &=& 0 , 
\end{eqnarray}
respectively, with
\begin{numcases}~ 
\mathcal{V}(r, R) = V(R+r/2) - V(R-r/2) , \label{eq: V-mat} \\
\mathcal{Q}(r, R, t) = - \frac{\hb^2}{m} \frac{1}{A} A_{rR} . \label{eq: Q-mat}
\end{numcases}
Here we have introduced the shorthand notations $A_r = \pa A / \pa r$ and $A_{rR} = \pa^2 A / \pa r \pa R$, with similar definitions for any other partial derivative. 
Equations \eqref{eq: HJ} and \eqref{eq: con} can be identified as the generalized  HJ and continuity equations in the present framework. In contrast to the standard closed-system case, the field $\mathcal{Q}$ in Eq. \eqref{eq: Q-mat} should be interpreted more carefully. Although it formally resembles the usual Bohmian quantum potential, it depends on the amplitude $A$ whose time evolution is governed by Eq.~\eqref{eq: con} and is therefore explicitly influenced by the thermal environment through the diffusion term. As a consequence, $\mathcal{Q}$ incorporates not only genuinely quantum contributions but also indirect thermal effects.
For this reason, it is more appropriate to interpret $\mathcal{Q}$ as an effective potential rather than a purely quantum one. In general, it displays a nontrivial interplay between quantum and environment-induced effects, since the $A$ amplitude shows
an explicit dependence on the temperature. As a consequence, the resulting contribution to the dynamics does not admit, in general, a simple separation into purely quantum and purely temperature-dependent parts, and may retain a residual dependence on temperature even in the classical regime.
To make this statement more precise, we can formally decompose the effective potential as
\begin{eqnarray} \label{eq: eff-pot}
\mathcal{Q}(r, R, t) = \mathcal{Q}_{\mathrm{res}}(r, R, t) + \mathcal{Q}_{\mathrm{van}}(r, R, t),
\end{eqnarray}
where $\mathcal{Q}_{\mathrm{res}}$ denotes the part that remains finite in the classical regime, while $\mathcal{Q}_{\mathrm{van}}$ is defined as the contribution that vanishes in that limit, that is,
\begin{eqnarray} \label{eq: van-pot}
\mathcal{Q}_{\mathrm{van}}(r, R, t) = \mathcal{Q}(r, R, t) - \mathcal{Q}_{\mathrm{res}}(r, R, t).
\end{eqnarray}
It should be emphasized that $\mathcal{Q}_{\mathrm{van}}$, although vanishing in the classical regime, may still carry an implicit dependence on temperature through the amplitude. Thus, the term ``purely quantum'' is to be understood from an operational view; namely, the part disappearing in the classical limit, rather than as a contribution entirely independent of environmental effects.
This observation highlights an important distinction with the conventional Bohmian formulation of closed systems, where the quantum potential vanishes in the classical regime. In this open-system framework, however, the effective potential retains a nontrivial structure, reflecting the persistent interplay between quantum coherence and thermal fluctuations. As an illustrative example, the effective potential for the propagation of a single Gaussian wave packet is analyzed in next Section.

It is also useful to recall that the density operator is Hermitian, namely $ \rho(x, x', t) = \rho^*(x', x, t) $. From Eq.~\eqref{eq: rho-polar} this immediately leads to 
\begin{numcases}~
A(r, R, t) = A(-r, R, t)  , \label{eq: A-even} \\
S(r, R, t) = -S(-r, R, t) , \label{eq: S-odd}
\end{numcases}
implying that the amplitude of the density matrix is an even function of the relative coordinate $r$, while its phase is odd in this coordinate. Recall also that the diagonal elements of the density matrix ($r=0$) are interpreted as the probability density, whereas the off-diagonal elements encode quantum coherences.

To proceed further in this context, from the properties \eqref{eq: A-even} and \eqref{eq: S-odd}, diagonal elements of the continuity equation \eqref{eq: con} read
\begin{eqnarray} \label{eq: con-1}
\frac{\pa P(x, t)}{\pa t} + \frac{\pa J(x, t)}{\pa x} &=& 0,  
\end{eqnarray}
where we have introduced the probability density $P(x, t)$ and the probability current density $J(x, t)$, respectively, as
follows
\begin{numcases}~
P(x, t) = A(r, R, t)\big|_{r=0, R=x} , \\
J(x, t) = P(x, t) \frac{1}{m} \frac{\pa S(r, R, t)}{\pa r} \bigg|_{r=0, R=x} .
\end{numcases}
Thus, the {\em Bohmian-like} velocity field is defined as 
\begin{eqnarray} \label{eq: bohm-vel}
v(x, t) &=& \frac{J(x, t)}{P(x, t)} = \frac{1}{m} S_r \big|_{r=0, R=x} ,
\end{eqnarray}
from which, as we have carried out in the Appendix, one can obtain a Newtonian-like equation according to
\begin{eqnarray} \label{eq: Newton}
\ddot{x} &=& - 2 \gamma \dot{x} - \frac{1}{m} \frac{\partial V}{\partial x} - \frac{1}{m} \frac{\partial \mathcal{Q}}{\partial x} \bigg|_{x'=x} ,
\end{eqnarray}
where, as discussed above, the last term represents an effective contribution originating from the nontrivial behavior of $\mathcal{Q}$. In general, this effective potential incorporates both genuine quantum and environment-induced effects of the amplitude in an inseparable manner. The resulting residual part displays a  dependence on the thermal environment, even in the classical regime. Explicitly, in the regime the Newtonian equation of motion one has that
\begin{eqnarray} \label{eq: class-Newton}
\ddot{x} &=& - 2 \gamma \dot{x} - \frac{1}{m} \frac{\partial V}{\partial x} - \frac{1}{m} \frac{\partial \mathcal{Q}_{\mathrm{res}}}{\partial x} \bigg|_{x'=x} ,
\end{eqnarray}
It is also instructive to compare Eq.~\eqref{eq: class-Newton} with the standard classical Langevin equation, which reads
\begin{eqnarray} \label{eq: langevin}
\ddot{x} = - 2 \gamma \dot{x} - \frac{1}{m} \frac{\partial V}{\partial x} + \frac{1}{m} \xi(t) ,
\end{eqnarray}
where $\xi(t)$ denotes a stochastic force characterizing thermal fluctuations and satisfying
\begin{eqnarray}
\langle \xi(t) \rangle = 0, \qquad
\langle \xi(t)\,\xi(t') \rangle \propto k_B T\, \delta(t-t').
\end{eqnarray}
A key critical difference between Eq.~\eqref{eq: class-Newton} and Eq.~\eqref{eq: langevin}  lies in the nature of thermal effects. In the Langevin framework, temperature enters through an explicit noise term, leading to intrinsically stochastic trajectories at the level of individual realizations. In contrast, in the present approach, thermal effects are fully encoded in the structure of the effective potential $\mathcal{Q}_{\mathrm{res}}$, which originates from the reduced density matrix dynamics.
As a consequence, trajectories generated by Eq.~\eqref{eq: class-Newton} remain deterministic, despite their explicit dependence on temperature. The effect of the environment is therefore not to introduce randomness in the dynamics of individual trajectories, but rather to modify the effective force field through which these trajectories evolve.
In this sense, the present formulation describes a class of temperature-dependent deterministic trajectories, which should be interpreted as representing an ensemble-encoded dynamics of the reduced density matrix, unlike the stochastic dynamics at the level of individual trajectories provided by the Langevin equation.

\subsection{Scaled Caldeira--Leggett equation in the quantum-to-classical transition}

Recently, a linear scaled equation has been proposed to describe the QCT within the von Neumann framework for isolated quantum systems \cite{MoMi-Sy-2023}. The main motivation behind this approach is to extend the quantum formalism, in particular the density matrix description, to classical systems. In this way, a nonlinear equation for the QCT is introduced and characterized by a quantumness parameter $\epsilon$. This parameter takes the value $\epsilon=1$ in the fully quantum regime, where the standard von Neumann equation is recovered, and reaches the classical regime at $\epsilon=0$, where the so-called classical von Neumann equation emerges. Interestingly enough, in-between regimes keep quantum features which are fading continuously and smoothly until reaching the classical regime.
The construction of this classical equation proceeds as follows. One first expresses the density matrix in polar form and substitutes it into the quantum von Neumann equation, which leads to a pair of coupled equations: a HJ equation and a continuity equation. In comparison with its classical counterpart, the corresponding HJ equation contains an additional term, known as the quantum potential, which accounts for quantum effects. By subtracting this term from the external potential and then reconstructing the dynamical equation in terms of the polar variables, one obtains an equation for the phase which is governed by the classical HJ equation. This motivates the interpretation of the resulting equation as a ``classical'' von Neumann equation. It can then be shown that, by introducing a scaling for both the density matrix and the Planck constant according to
\begin{eqnarray} \label{eq: hbt}
\tilde{\hbar} = \sqrt{\epsilon}\,\hbar ,
\end{eqnarray}
the nonlinear QCT equation becomes equivalent to a linear scaled equation for all non-classical regimes, including the fully quantum one. Despite the structural differences between the nonlinear and linear formulations at the level of the dynamical equations, a smooth transition is recovered  at the level of solutions
: starting from the solution of the scaled equation, the classical behavior is recovered continuously and smoothly by taking the limit $\epsilon \to 0$.

It should be emphasized that the quantumness parameter $\epsilon$ is not an experimentally measurable quantity or an independent physical parameter of the system. Rather, it is introduced as a theoretical interpolation parameter that continuously scales the contribution of the quantum potential, thereby providing a convenient framework for studying the crossover from the fully quantum description to the corresponding classical dynamics.

In the standard formulation of open quantum systems, the combined system consisting of the system of interest and its environment is treated as an isolated entity obeying the von Neumann equation. The reduced dynamics of the system is then obtained by tracing out the environmental degrees of freedom. Within the CL approach, the environment is modeled as a harmonic bath, and by employing the Born and Markov approximations together with an Ohmic spectral density, one arrives at the CL equation \eqref{eq: CLeq_Rr} in the high-temperature limit.
Following the same procedure for the scaled von Neumann equation of motion, one obtains a scaled version of the CL equation,
\begin{eqnarray} \label{eq: scaled-CLeq_Rr} 
\left[ \frac{\pa}{\pa t} - i \frac{\hbt}{m} \frac{\pa^2}{\pa R \pa r} + \frac{ V(R+r/2) - V(R-r/2) }{ i\hbt } + 2 \ga r \frac{\pa}{\pa r} + \frac{D}{\hbt^2} r^2 \right] \rhot(r, R, t) &=& 0 , 
\end{eqnarray} %
in terms of the scaled density matrix,  $ \rhot $,  and the scaled Planck constant  defined by Eq. \eqref{eq: hbt}. This equation has been recently used in the context of surface diffusion \cite{Miret2026}.

As discussed above, the scaled linear CL equation does not provide a proper description of the classical regime. In particular, the limit toward classicality is not obtained in a straightforward manner at the level of the dynamical equation. This suggests that the classical regime should instead be described by a different, inherently nonlinear equation.
The guiding principle for constructing such an equation is that its phase dynamics should reproduce the classical HJ equation, while still retaining the influence of the environment. From the analysis of the effective potential, we have seen that, in the classical regime, only its residual part contributes to the dynamics. This observation motivates a modification of the original CL equation in which the vanishing part of the effective potential is removed from the external potential term.
Accordingly, we propose the following ``classical'' CL equation:
\begin{eqnarray} \label{eq: class-CL}
\left[ \frac{\partial}{\partial t}
- i \frac{\hbar}{m} \frac{\partial^2}{\partial R \partial r}
+ \frac{ \mathcal{V} - \mathcal{Q}_{\mathrm{van}} }{ i\hbar }
+ 2 \gamma r \frac{\partial}{\partial r}
+ \frac{D}{\hbar^2} r^2 \right] \rho_{\mathrm{cl}}(r, R, t) = 0 ,
\end{eqnarray}
where $\mathcal{Q}_{\mathrm{van}}$ denotes the part of the effective potential that vanishes in the classical regime.

Proceeding as before, we substitute the polar form $\rho_{\mathrm{cl}} = A_{\mathrm{cl}} e^{i S_{\mathrm{cl}}/\hbar}$ into Eq.~\eqref{eq: class-CL}. This again leads to two coupled differential equations for the amplitude and phase. The phase equation reduces to a generalized Hamilton--Jacobi equation of the form
\begin{eqnarray} \label{eq: clas-HJ} 
- \frac{\partial S_{\mathrm{cl}}}{\partial t}
&=& \frac{1}{m} \frac{\partial S_{\mathrm{cl}}}{\partial r} \frac{\partial S_{\mathrm{cl}}}{\partial R}
+ 2 \gamma r \frac{\partial S_{\mathrm{cl}}}{\partial r}
+ \mathcal{V} + \mathcal{Q}_{\mathrm{res}} ,
\end{eqnarray}
from which the corresponding classical equation of motion \eqref{eq: class-Newton} follows. In this way, the proposed equation consistently captures the classical dynamics while retaining the residual, environment-induced contribution to the effective potential.

\section{Dynamics of Gaussian wave packets}

In this section, we study the free evolution of Gaussian wave packets. We begin with a single Gaussian wave packet and then extend the analysis to a superposition of two such states.

\subsection{A single Gaussian wave packet}

Let us first consider an initial state described by the scaled Gaussian wave packet
\begin{eqnarray} \label{eq: wf0}
\widetilde{\psi}_0(x) &=& \frac{1}{(2\pi \si_0^2)^{1/4}} \exp \left[ - \frac{(x-x_0)^2}{4\si_0^2} + i \frac{p_0}{\hbt} x \right],
\end{eqnarray}
where $x_0$, $p_0$, and $\sigma_0$ denote the initial position of the center, the initial momentum, and the initial width, respectively.

To solve Eq.~\eqref{eq: scaled-CLeq_Rr}, we proceed in three steps. First, a partial Fourier transform with respect to the center-of-mass coordinate is performed. The resulting equation is then solved, and finally the inverse transform is applied to recover the density matrix in the position representation \cite{Ve-PRA-1994, VeKuGh-PA-1995}. Following this procedure, we obtain
\begin{eqnarray}
\rhot(r, R, t) &=& \frac{1}{ \sqrt{2\pi} \wt_t } \exp\left[ \bzet(r, t) - \frac{ ( R - \bont(r, t) )^2 }{ 2 \wt_t^2} \right] \label{eq: denmat}  \\
\jt(r, R, t) &=& -i\frac{\hbt}{m} \left( \frac{\pa \bzet}{\pa r} + \frac{ x-\bont(0, t) }{\wt_t^2} \frac{\pa \bont}{\pa r} \right) \rho(r, R, t) \label{eq: curmat}
\end{eqnarray}
for the density matrix and the associated current density matrix. The functions $\bzet(r,t)$ and $\bont(r,t)$ are given by
\begin{eqnarray}
\bzet(r, t)& =& - \left[ \frac{ e^{-4\ga t} }{ 8 \si_0^2 } + \frac{ 1 - e^{-4\ga t} }{4\ga} \frac{D}{\hbt^2} \right]r^2
+ i \left( \frac{p_0}{\hbt} ~e^{-2\ga t}  \right) r
\label{eq: b0}
	\\
\bont(r, t) &=& x_t  
+ i  \left[ \frac{ \hbt }{ 4 m \si_0^2 } e^{-2\ga t} \uptau(t) + \frac{D}{m \hbt} \uptau(t)^2 \right] r
\label{eq: b1} \\ 
\wt_t &=& \si_0 \sqrt{1 + \frac{\hbt^2 }{ 4 m^2 \si_0^4 } \uptau(t)^2
+ \frac{ 4\ga t + 4 e^{-2\ga t} - 3 - e^{-4\ga t} }{8 m^2 \ga^3 \si_0^2} ~ D }  
\label{eq: wt}
\end{eqnarray}
with
\begin{eqnarray} \label{eq: tau}
\uptau(t) &=& \frac{1-e^{-2\ga t}}{2\ga} ,
\end{eqnarray}
and
\begin{eqnarray}\label{eq: xt}
x_t &=& x_0 + \frac{p_0}{m} \uptau(t) .
\end{eqnarray}
Note that $x_t$ is the solution of the classical equation of motion \eqref{eq: class-Newton} subject to the initial condition $x(0)=x_0$. For this reason, we refer to $x_t$ as the classical path. As will be shown below, the last term in Eq.~\eqref{eq: class-Newton} vanishes identically along this trajectory, so that the equation effectively reduces to its form without that contribution. 

From Eq. \eqref{eq: denmat}, the corresponding scaled effective potential, having the same functional form as \eqref{eq: Q-mat} but expressed in terms of scaled quantities, is obtained as
\begin{eqnarray}
\Qefft(r, R, t) &=& - \frac{\tilde{n}(r, R, t)}
{\left(2 D \si_0^2 \left(e^{4 \ga  t} (4 \ga  t-3)+4 e^{2 \ga  t}-1\right) + 16 \ga ^3 m^2 \si_0^4 e^{4 \ga  t} + \ga   \hbt^2 \left(e^{2 \ga  t}-1\right)^2\right)^2 }
\end{eqnarray}
where the numerator is defined as
\begin{eqnarray}
\tilde{n}(r, R, t) &=& 4 \ga  \si_0^2 e^{2 \ga  t} \left(p_0-e^{2 \ga  t} (2 \ga  m (x_0-R)+p_0)\right)
\times \nonumber \\
&&
\bigg[
32 D^2 \si_0^2 e^{2 \ga  t} \sinh (\ga  t) (\ga  t \cosh (\ga  t)-\sinh (\ga  t)) + 16 \ga ^3 D m^2 \si_0^4 \left(e^{4 \ga  t}-1\right) 
\nonumber \\
&+&
\hbt^2 \bigg( \ga  D \left( 4 \ga  t-4 e^{2 \ga  t}+e^{4 \ga  t}+3 \right) + 8 \ga^4 m^2 \si_0^2 \bigg) \bigg] ~ r .
\end{eqnarray}

In the classical regime $\ep = 0$, the effective potential reduces to
\begin{eqnarray} \label{eq: Qres-gauss}
\mathcal{Q}_{\res}(r, R, t) &=& - \frac{n_{\res}(r, R, t)}
{\left(2 D \si_0^2 \left(e^{4 \ga  t} (4 \ga  t-3) + 4 e^{2 \ga  t}-1\right) + 16 \ga ^3 m^2 \si_0^4 e^{4 \ga  t} \right)^2 }
\end{eqnarray}
which represents the residual contribution that survives in this regime, with
\begin{eqnarray}
n_{\res}(r, R, t) &=& 4 \ga \si_0^2 e^{2 \ga  t} \left(p_0-e^{2 \ga  t} (2 \ga  m (x_0-R)+p_0)\right) 
\times \nonumber \\
&&
\bigg(
32 D^2 \si_0^2 e^{2 \ga  t} \sinh (\ga  t) (\ga  t \cosh (\ga  t)-\sinh (\ga  t))+ 16 \ga ^3 D m^2 \si_0^4 \left(e^{4 \ga  t}-1\right)
\bigg)  ~ r .
\end{eqnarray}

By setting $r=0$, the scaled probability density (PD) and probability current density (PCD) are obtained as
\begin{eqnarray}
\Pt(x, t)& =& \frac{1}{ \sqrt{2\pi} \wt_t } \exp\left[ - \frac{ ( x - x_t )^2 }{ 2 \wt_t^2} \right] , \label{eq: probden}  \\
\tilde{J}(x, t) &=& \left[ \frac{p_0}{m} e^{-2\ga t} + \frac{x-x_t}{\wt_t^2} \tau(t) 
\left(\frac{\hbt^2}{4m^2\si_0^2} e^{-2\ga t} + \frac{D}{m^2} \tau(t) \right) \right] \Pt(x, t) .
\label{eq: cur}
\end{eqnarray}
The probability density thus retains its Gaussian form with a time-dependent width $\wt_t$, while its center follows the classical trajectory $x_t$.

From Eq. \eqref{eq: Qres-gauss}, one now obtains the residual force defined as
\begin{eqnarray} \label{eq: res-force}
F_{\mathrm{res}}(x, t) &=& 
 - \frac{1}{m} \frac{\partial \mathcal{Q}_{\mathrm{res}}}{\partial x} \bigg|_{x'=x} =
 - \frac{1}{m} \frac{\partial \mathcal{Q}_{\mathrm{res}}}{\partial r} \bigg|_{R=x, r=0} 
\nonumber \\
&=& \frac{64 \ga^2 D m e^{6 \ga  t} \sinh (\ga  t) \left[ \ga  \cosh (\ga  t) \left(D t+2 \ga ^2 m^2 \si_0^2\right)-D \sinh (\ga  t)\right] }
{\left( D - 4 D e^{2 \ga  t} + D e^{4 \ga  t} ( 3-4 \ga t ) -8  \ga^3 m^2 \si_0^2 e^{4 \ga  t} \right)^2 }
~(x-x_t) ,
\end{eqnarray}
appearing in the classical equation of motion \eqref{eq: class-Newton}.

The structure of the dynamics reveals a clear distinction between dissipation and thermal effects. The trajectory $x_t$ depends explicitly on the damping coefficient $\ga$, reflecting the role of friction, while it remains insensitive to temperature. This follows from the fact that the residual force vanishes along the trajectory $ x_t $. 
At first sight, this may appear counterintuitive, since thermal fluctuations are often associated with random forces acting directly on trajectories. In the present framework, however, thermal effects do not manifest as stochastic forces at the level of individual trajectories. Instead, they enter through the effective potential and influence the dynamics indirectly via the deformation of the probability distribution.
As a result, trajectories remain deterministic in all regimes, including the classical one, while still exhibiting a nontrivial dependence on temperature. In particular, trajectories corresponding to initial positions away from the center ($x \neq x_0$) are affected by temperature through the residual force \eqref{eq: res-force}, which encodes the environment-induced contribution to the dynamics.

In the free-friction limit, where the damping term (i.e., the second term on the right-hand side of Eq.~(\ref{eq: CL eq})) is neglected, Eqs.~(\ref{eq: xt}) and (\ref{eq: wt}) reduce to
\begin{numcases}~
x_t \approx x_0 + \frac{p_0}{m} t , \label{eq: xt_limit} \\
\wt_t \approx \si_0 \sqrt{1 +  \frac{ \hbt^2 }{ 4 m^2 \si_0^4 } t^2 + \frac {2 D }{ 3 m^2 \si_0^2 } ~ t^3 }  . \label{eq: wt_limit}
\end{numcases}

The $\ell_1$-norm of quantum coherence, which is directly related to the coherence length (i.e., the width of the density matrix along the off-diagonal direction $x'=-x$, or equivalently $r=2x$), is defined for continuous-variable systems as
\begin{equation}
\widetilde{C}_{\ell_1}(t) = \int dr \int dR \, \big| \rhot(r, R, t) \big| .
\end{equation}
In the present case, this quantity can be evaluated explicitly as
\begin{eqnarray} \label{eq: cl1}
\widetilde{C}_{\ell_1}(t) &=& \sigma_0 \sqrt{\pi}
\left[ \frac{ e^{-4\gamma t} }{ 8 } + \frac{ 1 - e^{-4\gamma t} }{4\gamma} \frac{D \sigma_0^2}{\hbt^2} \right]^{-1/2}
\nonumber \\
&& \times \exp \left[ -\frac{2 \gamma p_0^2 \sigma_0^2}{ 2 D \sigma_0^2 \left(e^{4 \gamma t}-1 \right) + \gamma \hbt^2 } \right] .
\end{eqnarray}
Its initial, short-time ($\gamma t \ll 1$) and long-time (stationary) values are given by
\begin{eqnarray} 
\widetilde{C}_{\ell_1}(0) &=& \sigma_0 \sqrt{8 \pi} 
\exp \left[ -\frac{2 p_0^2 \sigma_0^2 }{ \hbt^2 } \right], \label{eq: cl1-0} \\
\widetilde{C}_{\ell_1}(t)
& \simeq &
\widetilde{C}_{\ell_1}(0)
\left[
1
+
t \left(
\frac{16 D p_0^2 \sigma_0^4}{\hbt^4}
- 4\frac{D\sigma_0^2}{\hbt^2}
+ 2\gamma
\right)
\right], \label{eq: cl1-shottime} \\
\widetilde{C}_{\ell_1}(t) \big|_{t \to \infty} &=&  \sqrt{ \frac{2\pi \hbt^2}{m k_B T} } . \label{eq: cl1-st}
\end{eqnarray}
The stationary value is just the scaled thermal wavelength. This behavior shows that, within a given regime, coherence decreases with increasing temperature, while for a fixed temperature it vanishes in the classical limit $\epsilon \to 0$.
For comparison, in the absence of environmental effects, i.e., within the scaled Schr\"odinger framework, one finds
\begin{eqnarray} \label{eq: cl1-Sch}
\widetilde{C}_{\ell_1, \mathrm{Sch}}(t) &=& \sigma_0 \sqrt{8 \pi}
\exp \left[ -\frac{2 p_0^2 \sigma_0^2 }{ \hbt^2 } \right],
\end{eqnarray}
which remains constant in time. This highlights the role of the environment in driving the decay of coherence in the present open-system dynamics.

\subsection{The Schr\"{o}dinger cat state}

We now turn to the case where the initial state is given by a coherent superposition of two wave packets,
\begin{eqnarray} \label{eq: sup0}
\psit_0(x) &=& \Nt ( \psit_{0a}(x) + \psit_{0b}(x) ) 
\end{eqnarray}
with $ \Nt $ the normalization constant. This state represents a prototypical Schr\"{o}dinger cat state, in which two distinct components coexist and give rise to interference effects.
From Eq.~(\ref{eq: sup0}), the initial density matrix takes the form
\begin{eqnarray} \label{eq: rho0}
\rhot(x, x', 0) &=& \Nt^2 ( \rhot_{aa}(x, x', 0) + \rhot_{ab}(x, x', 0) + \rhot_{ba}(x, x', 0) + \rhot_{bb}(x, x', 0) ) , 
\end{eqnarray}
where $\rhot_{ij}(x,x',0)=\psit_{0i}(x)\psit_{0j}^*(x')$, with $i,j=a,b$. The diagonal terms describe the individual wave packets, while the off-diagonal terms encode the coherence between them and are therefore responsible for interference.

Due to the linearity of the master equation~(\ref{eq:  scaled-CLeq_Rr}) with $r=0$, each component of Eq.~(\ref{eq: rho0}) evolves independently. The full density matrix is then obtained by superposing these evolved terms. As a result, the PD can be constructed as
\begin{eqnarray} \label{eq: probden_sup}
\Pt(x, t) &=& \Nt^2 ( \Pt_{aa}(x, t) + \Pt_{ab}(x, t) + \Pt_{ba}(x, t) + \Pt_{bb}(x, t) )   .
\end{eqnarray}
By using the relation $\Pt_{ba}(x,t)=\Pt_{ab}^*(x,t)$, the PD can be written as
\begin{eqnarray} \label{eq: probden_sup1}
\Pt(x, t) &=& \Nt^2 ( \Pt_{aa}(x, t) + \Pt_{bb}(x, t) + 2 |\Pt_{ab}(x, t)| \cos [\theta(x, t)] ) ,
\end{eqnarray}
where $|\Pt_{ab}(x,t)|$ and $\theta(x,t)$ denote the magnitude and phase of the interference term, respectively. Rewriting this expression in the standard interference form \cite{BaPE-book-2002},
\begin{eqnarray} \label{eq: probden_sup2}
\Pt(x, t) &=& \Nt^2 \left( \Pt_{aa}(x, t) + \Pt_{bb}(x, t) + 2 \sqrt{ \Pt_{aa}(x, t) \Pt_{bb}(x, t) } ~ e^{\Gat(t)} 
\cos [\tilde{\theta}(x, t)] \right)   ,
\end{eqnarray}
which makes explicit that the interference pattern is governed by the competition between the diagonal contributions and the coherence encoded in the off-diagonal terms. The time-dependent function
\begin{eqnarray} \label{eq: dech-measure}
\Gat(t) &=& \log \frac{|\Pt_{ab}(x, t)|}{ \sqrt{ \Pt_{aa}(x, t) \Pt_{bb}(x, t) } } 
\end{eqnarray}
serves as a natural measure of decoherence. It quantifies the relative weight of the interference term with respect to the classical mixture. A negative value of $\Gat(t)$ points out suppression of coherence, i.e., the progressive disappearance of interference fringes. Furthermore, the so-called coherence attenuation function is defined as \cite{FC-PLA-2001}
\begin{eqnarray} \label{eq: att-coef}
\tilde{a}(t) &=& e^{\Gat(t)} ,
\end{eqnarray}
which provides a direct measure of how fast coherence is lost due to the interaction with the environment.

Let us now consider the physically relevant case of two Gaussian wave packets initially localized symmetrically around the origin, with equal widths and opposite momenta,
\begin{eqnarray} \label{eq: wf0_sup} 
\psit_0(x) &=& \Nt \frac{1}{(2\pi \si_0^2)^{1/4}} \left\{ \exp \left[ - \frac{(x-x_0)^2}{4\si_0^2} + i \frac{p_0}{\hbt} x \right] + \exp \left[ - \frac{(x+x_0)^2}{4\si_0^2} - i \frac{p_0}{\hbt} x \right] \right\} 
\end{eqnarray}
with normalization constant
\begin{eqnarray} \label{eq: nor}
\Nt &=& \left\{ 2 + 2 \exp \left[ - \frac{x_0^2}{2\si_0^2} - \frac{2 p_0^2 \si_0^2}{\hbt^2}  \right] \right\}^{-1/2} .
\end{eqnarray}

In this case, one obtains
\begin{numcases}~ 
\Gat(t) = -\left( \frac{x_0^2}{2\si_0^2} + 2 \frac{p_0^2}{\hbt^2} \si_0^2 \right)   
\left[ 1 - \frac{\si_0^2}{\wt_t^2} \left( 1 +  \frac{\hbt^2}{4m^2\si_0^4} \uptau(t)^2 \right) \right ]   , \label{eq: deco_func}
	\\
\tilde{\theta}(x, t) = \frac{ \tilde{\beta}(t)}{\wt_t^2} x    , \label{eq: phase}
\end{numcases}
for the decoherence function and the phase, respectively, where
\begin{eqnarray}
\tilde{\beta}(t) &=& - x_0 \frac{\hbt}{2m\si_0^2} \uptau(t) - 2 \frac{p_0}{\hbt} \si_0^2 \label{eq: beta}   
\end{eqnarray}
and the temperature enters through the diffusion coefficient. 
The structure of the decoherence function shows that environmental effects systematically reduce the weight of the interference term. In particular, the relevant contribution is always such that $\Gamma(t)$ remains negative, reflecting the continuous suppression of coherence over time.
As expected, in the absence of environment, the decoherence function vanishes. In this limit, the Caldeira--Leggett equation reduces to the standard Schr\"{o}dinger equation, and the superposition remains perfectly coherent. This highlights that decoherence is entirely induced by the coupling to the environment.
Finally, in the classical regime, corresponding to the limit $\ep \to 0$, the decoherence function diverges to $-\infty$. This results in a vanishing attenuation parameter, thereby producing a purely classical mixture of the two wave packets, as described by Eq. \eqref{eq: probden_sup2}. In this regime, interference effects disappear completely, and the system behaves as an incoherent statistical ensemble rather than a coherent superposition.

\section{Results and discussion}

In this section, we present our numerical results and discuss their physical implications. Throughout the calculations, we set $\hb = k_B = m = 1$. The initial Gaussian wave packet is characterized by the parameters $x_0=-5$, $p_0=2$, and $\si_0=1$. Unless otherwise stated, the relaxation rate is fixed at $\ga = 0.01$.

\begin{figure} 
	\centering
	\includegraphics[width=12cm,angle=-0]{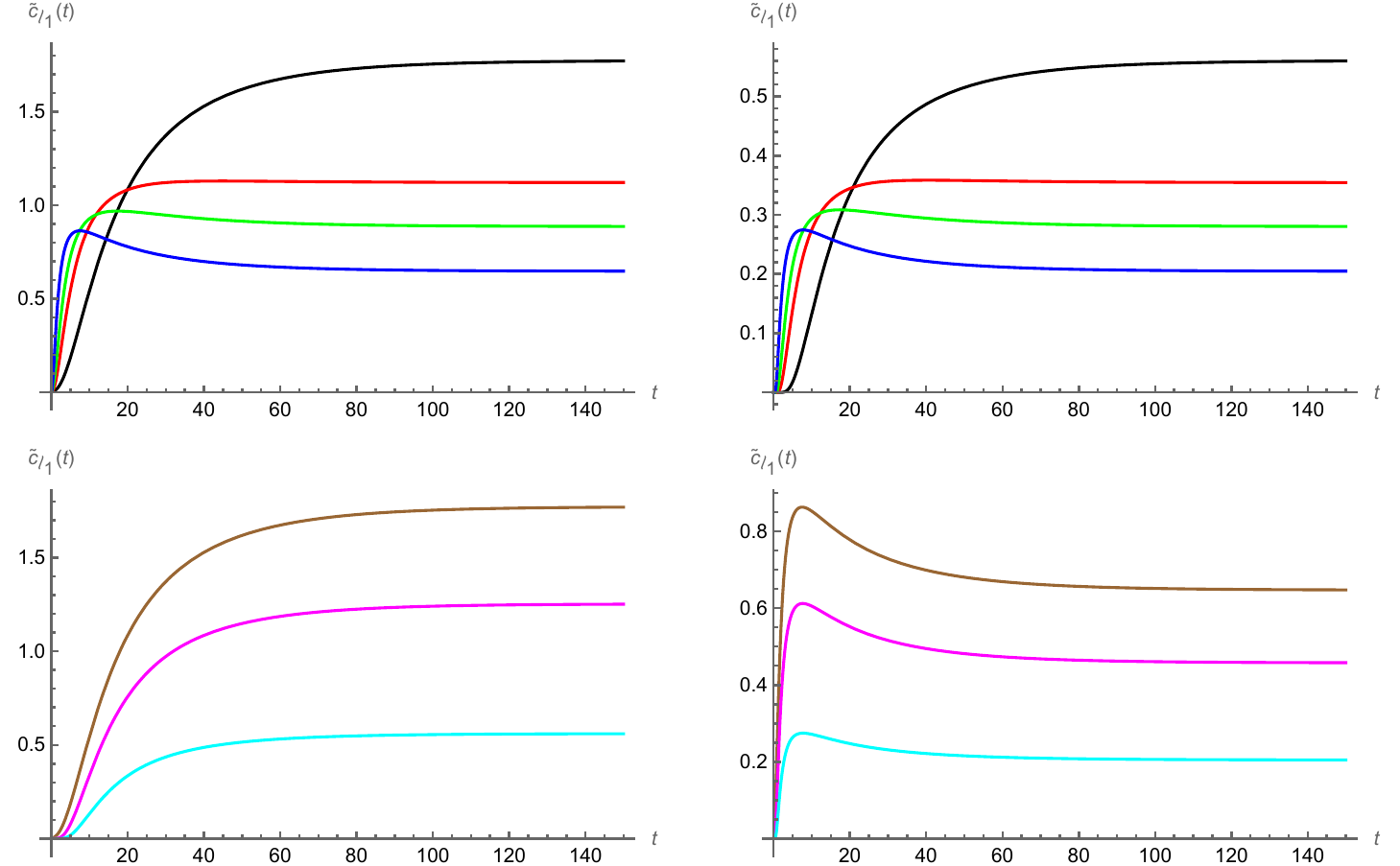}
	\caption{Quantum coherence given by \eqref{eq: cl1}. Top panels: quantum regime (left) and transition regime with $\ep=0.1$ (right) for different temperatures: $T=2$ (black), $T=5$ (red), $T=8$ (green), and $T=15$ (blue). Bottom panels: coherence for fixed temperatures $T=2$ (left) and $T=15$ (right), for different regimes $\ep=1$ (brown), $\ep=0.5$ (magenta), and $\ep=0.1$ (cyan). Other parameters are $\hb = m = k_B = 1$, $x_0=-5$, $p_0=2$, $\si_0=1$, and $\ga=0.01$.}
	\label{fig: quantum_coh} 
\end{figure}

Figure~\ref{fig: quantum_coh} displays the time evolution of the $\ell_1$-norm of coherence $\widetilde{C}_{\ell_1}(t)$ given in Eq.~\eqref{eq: cl1}. Several features of the dynamics can be directly understood from the analytical structure of this expression.
Let us first consider the top panels, where different temperatures are compared. In both the quantum regime ($\epsilon=1$) and the transition regime ($\epsilon=0.1$), the coherence evolves from its initial value toward a stationary value. This saturation is fully consistent with the long-time limit \eqref{eq: cl1-st}, which depends explicitly on both the temperature and the parameter $\epsilon$.
The effect of temperature is clearly visible: increasing $T$ reduces the stationary value of coherence. This follows directly from the inverse square-root dependence in Eq.~\eqref{eq: cl1-st}, and reflects the expected physical behavior that stronger thermal fluctuations suppress quantum coherence more efficiently.
Immediately after $t=0$, the coherence corresponding to higher temperatures exceeds that obtained at lower temperatures. However, this does not originate from different initial coherence values, since Eq.~\eqref{eq: cl1-0} shows that $\widetilde{C}_{\ell_1}(0)$ is independent of temperature. Instead, it results from the temperature dependence of the initial evolution rate. According to the short-time expansion in Eq.~\eqref{eq: cl1-shottime}, the initial slope is proportional to the diffusion coefficient $D=2\gamma k_B T$, so that, for the parameters considered here, the initial growth becomes steeper as the temperature increases. In the transition regime, this effect is further amplified because $\widetilde{\hbar}=\sqrt{\epsilon}\hbar$, leading to contributions that scale as $1/\epsilon$ and $1/\epsilon^2$. Nevertheless, this transient behavior does not indicate weaker thermal decoherence; at longer times thermal diffusion dominates the dynamics, yielding lower stationary coherence values for higher temperatures.
In the quantum regime (top left), the coherence therefore exhibits a rapid initial increase followed by a smooth relaxation toward its asymptotic value. As the temperature increases, the asymptotic coherence decreases and the relaxation proceeds over a shorter timescale. In the transition regime (top right), the same qualitative behavior is observed, but the coherence remains substantially suppressed because of the reduced effective Planck constant. Although the initial growth is enhanced through the $1/\epsilon$- and $1/\epsilon^2$-dependent terms in Eq.~\eqref{eq: cl1-shottime}, the overall coherence stays significantly smaller over an extended time interval.
The bottom panels isolate the role of $\epsilon$ by fixing the temperature. In this case, all curves approach stationary values that differ only by an overall $\sqrt{\epsilon}$ factor, as dictated by Eq.~\eqref{eq: cl1-st}. In other words, at a given temperature, the stationary coherence decreases proportionally to $\sqrt{\epsilon}$. This shows that while thermal fluctuations set the characteristic coherence scale, the parameter $\epsilon$ controls its overall magnitude, suppressing coherence as the system moves toward the classical limit.
A more subtle feature appears when comparing low and high temperatures. At low temperature (bottom left), the growth of coherence is essentially monotonic. At higher temperature (bottom right), a slight overshoot can occur before relaxation. This behavior can be qualitatively understood from the interplay between different diffusion-driven contributions in Eq.~\eqref{eq: cl1}, which affect the exponential damping and the normalization in distinct ways. At higher temperatures, the diffusion-induced terms grow more rapidly, modifying the balance between these contributions and leading to a transient enhancement of coherence before the eventual relaxation toward the stationary value. (That said, this overshoot is a secondary effect and depends sensitively on the parameters.)
%

\begin{figure} 
	\centering
	\includegraphics[width=12cm,angle=-0]{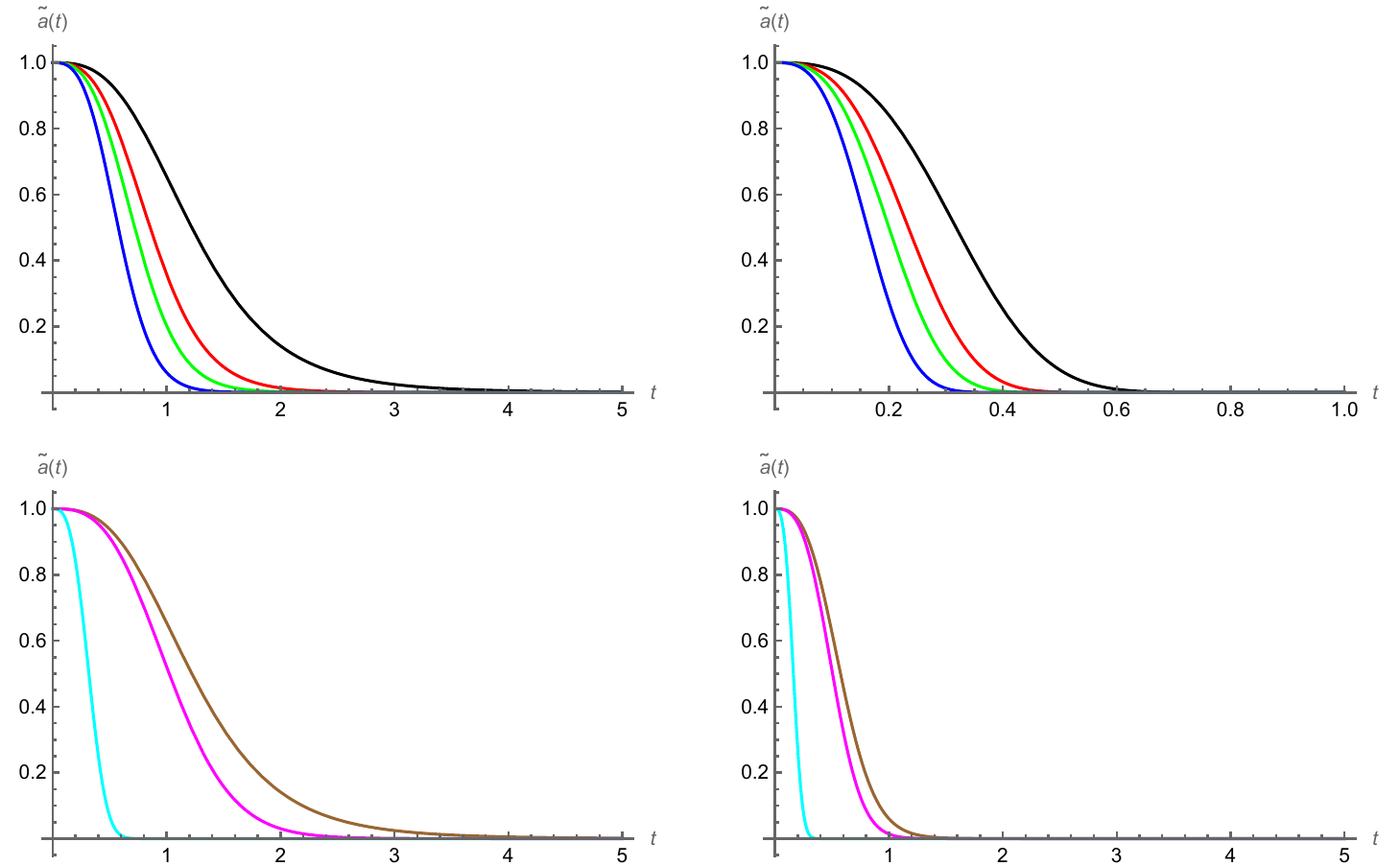}
	\caption{Coherence attenuation coefficient $\tilde{a}(t)$ given by \eqref{eq: att-coef} versus time. Top panels show the quantum regime (left) and the transition regime with $\ep=0.01$ (right) for different temperatures: $T=2$ (black), $T=5$ (red), $T=8$ (green), and $T=15$ (blue). Bottom panels show $a(t)$ for fixed temperatures $T=2$ (left) and $T=15$ (right) for different regimes: $\ep=1$ (brown), $\ep=0.5$ (magenta), and $\ep=0.01$ (cyan). Other parameters are fixed as $\hb=m=k_B=1$, $x_0=-5$, $p_0=2$, $\si_0=1$, and $\ga=0.01$. These plots highlight how both temperature and quantumness influence the rate of coherence loss.}
	\label{fig: attenuation_func} 
\end{figure}

Figure~\ref{fig: attenuation_func} shows the time dependence of the coherence attenuation coefficient $\tilde{a}(t)$ defined in Eq.~\eqref{eq: att-coef}, for the superposition state \eqref{eq: wf0_sup}. Since $\tilde{a}(t)=e^{\Gat(t)}$, it directly reflects the decay of the interference term relative to the classical contributions. Thus, the decrease of $\tilde{a}(t)$ toward zero corresponds to the suppression of coherence and the disappearance of interference fringes.
Let us first consider the top panels, where different temperatures are shown. In both the quantum regime and the transition regime, $\tilde{a}(t)$ starts from unity and decays monotonically to zero. This indicates that the system initially possesses strong coherence, which is gradually destroyed by the interaction with the environment.
The role of temperature is clearly visible: higher temperatures lead to a faster decay of $\tilde{a}(t)$. This behavior is expected, since increasing $T$ enhances the diffusion coefficient and therefore strengthens the coupling to the environment. As a result, the interference term $\widetilde{P}_{ab}$ is suppressed more rapidly leading to a quicker loss of coherence.
Comparing the quantum regime (top left) with the transition regime (top right), one observes that the decay is generally faster in the transition case. This reflects the fact that reducing the quantumness parameter $\ep$ weakens the phase coherence between the two wave packets, making the interference term more fragile against environmental effects.
The bottom panels provide further insight by fixing the temperature and varying $\ep$. At low temperature (bottom left), the decay of $\tilde{a}(t)$ is relatively slow, and the dependence on $\ep$ is clearly visible: smaller values of $\ep$ lead to a faster loss of coherence. This shows that quantum coherence is more robust when the system is deeper in the quantum regime.
At higher temperature (bottom right), the situation changes qualitatively. The decay becomes very rapid for all values of $\ep$. In this regime, thermal fluctuations dominate the dynamics, and the effect of the quantumness parameter becomes less pronounced. In other words, the environment washes out the distinction between different regimes, driving the system quickly toward a classical mixture.
Overall, the figure demonstrates a clear interplay between temperature and quantumness. While both contribute to the suppression of coherence, their roles are not equivalent: the parameter $\ep$ controls the intrinsic ability of the system to sustain interference, whereas the temperature governs how strongly the environment acts to destroy it. In the long-time limit, $\tilde{a}(t) \to 0$ in all cases, confirming that the superposition state ultimately decoheres into a classical statistical mixture.

%

\begin{figure} 
	\centering
	\includegraphics[width=15cm,angle=-0]{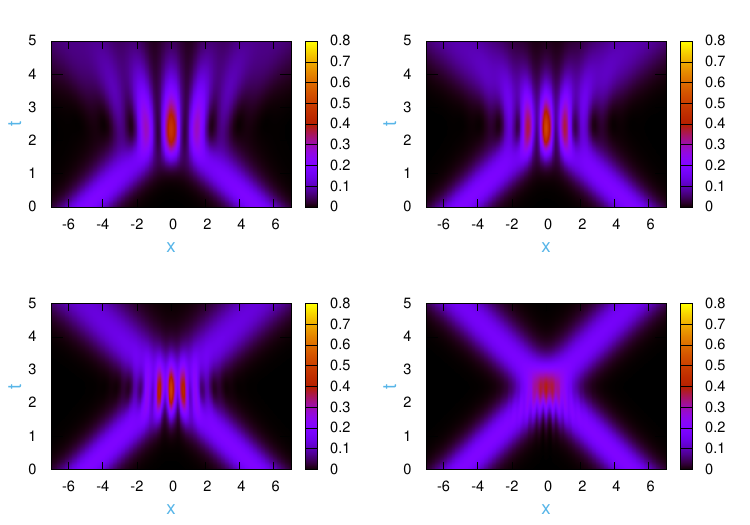}
	\caption{Density plots of the probability density for the superposition state \eqref{eq: wf0_sup} in the $x$-$t$ plane within the framework of the scaled CL equation for $\ga=0.001$ and $T=1$. Different panels correspond to quantumness parameters: $\ep=1$ (top left), $\ep=0.5$ (top right), $\ep=0.2$ (bottom left), and $\ep=0.05$ (bottom right). Other parameters are fixed as $\hb=m=k_B=1$, $x_0=-5$, $p_0=2$ and $\si_0=1$. These figures show how the environment gradually suppresses coherence as $\ep$ decreases.}
	\label{fig: denplotCL} 
\end{figure}

The density plots in Fig.~\ref{fig: denplotCL} provide a direct space--time visualization of how coherence is progressively degraded by the environment. What is particularly illuminating here is that the loss of coherence is not an abstract statement about matrix elements, but appears explicitly as the fading of interference fringes in the $x$--$t$ plane.
Let us begin with the top-left panel, corresponding to the quantum regime. One clearly observes pronounced interference fringes around the center, where the two counter-propagating Gaussian wave packets overlap. These fringes are the direct manifestation of the cross-term contribution $\widetilde{P}_{ab}(x,t)$ in Eq.~\eqref{eq: probden_sup1}. In this regime, coherence is still robust: the interference pattern persists over a relatively long time, indicating that $\Gat(t)$ remains close to zero and thus $\tilde{a}(t)$ stays near unity for a significant interval.
Moving to the top-right panel ($\ep=0.5$), the same qualitative structure is present, but the interference fringes are visibly weakened and become less extended in time. This signals that the relative weight of $\tilde{P}_{ab}$ is already being suppressed. In other words, $\Gat(t)$ becomes more negative at earlier times, and the attenuation function $a(t)$ decays faster. The environment is now more efficient in washing out phase relations between the two components of the superposition.
This tendency becomes much more pronounced in the bottom panels. For $\epsilon=0.2$, the interference fringes are confined to a shrinking overlap region in the $x$--$t$ plane: they are visible only near $x\approx 0$ and decay rapidly in time, reflecting the faster suppression of the off-diagonal contribution $\tilde{P}_{ab}(x,t)$. The dynamics starts to resemble that of a classical mixture: the two wave packets effectively propagate without maintaining a significant phase correlation. Finally, in the bottom-right panel ($\epsilon=0.05$), the interference structure is almost completely lost. What remains is essentially a single, smooth distribution with no visible oscillatory pattern, indicating that $\tilde{P}_{ab}$ has become negligible compared to the diagonal terms $\tilde{P}_{aa}$ and $\tilde{P}_{bb}$.
From a physical point of view, these plots clearly show that decreasing $\ep$ enhances the effectiveness of environmental decoherence. The parameter $\ep$ controls the degree of ``quantumness'' of the dynamics, and as it is reduced, the system becomes increasingly sensitive to the dissipative and noisy background. As a result, the interference term is rapidly damped, driving $\Gat(t)$ to large negative values and forcing $\tilde{a}(t)$ toward zero on shorter time scales.
In summary, Fig.~\ref{fig: denplotCL} offers a transparent picture of decoherence in action: the gradual disappearance of interference fringes directly reflects the decay of $\tilde{a}(t)$ and the suppression of quantum coherence. The transition from a fully quantum regime to an effectively classical mixture is thus seen not as an abrupt change, but as a continuous dynamical process controlled by $\epsilon$ and mediated by the environment.

\section{Summary and conclusions}

In this work, we have developed a unified framework for dissipative quantum dynamics across the quantum-classical boundary within the Caldeira-Leggett setting. The starting point of our construction is the observation that, in the standard formulation of open quantum systems, the combined system consisting of the system of interest and its environment is described as an isolated entity obeying the von Neumann equation. The reduced dynamics is then obtained by tracing out the environmental degrees of freedom. Within the Caldeira--Leggett approach, where the environment is modeled as a harmonic bath, the use of the Born and Markov approximations together with an Ohmic spectral density leads, in the high-temperature limit, to the standard Caldeira-Leggett equation \eqref{eq: CLeq_Rr}.

Following the same reduction procedure, but now guided by a recently proposed scaled von Neumann framework for isolated quantum systems, we anticipate a corresponding scaled version of the CL equation in terms of a scaled density matrix and a scaled Planck constant. In this formulation, both the density matrix and Planck constant are rescaled through a quantumness parameter, which interpolates continuously between the quantum and classical regimes. This provides a consistent embedding of dissipative dynamics into a broader quantum-classical transition framework.

We further showed that the classical regime cannot, in general, be obtained by a straightforward limiting procedure applied to the scaled linear equation. Instead, consistency with the Hamilton--Jacobi structure requires a modified description in which the dynamics becomes effectively nonlinear. Motivated by this observation, we introduced a classical Caldeira--Leggett equation in which the non-vanishing contribution of the effective potential is appropriately removed from the external potential term, ensuring that the phase dynamics reduces to the classical Hamilton--Jacobi equation while still retaining the influence of the environment through a residual potential.

A central element of the present formulation is the effective potential, which encodes quantum effects and environmental contributions in a nontrivial and generally inseparable manner. Importantly, this potential does not vanish in the classical regime and cannot, in general, be decomposed into independent quantum and thermal parts. Its residual contribution survives in the classical limit and modifies the resulting dynamics in a nontrivial way.

We then analyzed the structure of trajectories emerging from this framework. The associated trajectories remain deterministic in all regimes, including the classical one, despite the presence of thermal fluctuations. In contrast to the classical Langevin description, where thermal noise is implemented through stochastic forces at the level of individual realizations, here temperature enters only through the effective potential and thus affects the probability distribution rather than introducing randomness at the trajectory level.

To illustrate these features, we studied the propagation of a Gaussian wave packet. While the center follows a dissipative trajectory governed by friction, thermal effects manifest through the spreading of the wave packet and the structure of the effective potential. In particular, trajectories away from the center exhibit a nontrivial temperature dependence through the residual force, highlighting the indirect but persistent role of the environment.

At the level of coherence, we first analyzed a single Gaussian wave packet using the $\ell_1$-norm, showing how the coherence length is progressively reduced by environmental coupling as well as by the decrease of the quantumness parameter $\epsilon$ toward the classical regime. We then considered a superposition of two counter-propagating Gaussian wave packets, where decoherence is directly observed through the attenuation of interference fringes. The results consistently demonstrate a gradual suppression of coherence, driven jointly by thermal effects and the approach to the classical regime.

The present analysis is restricted to the high-temperature Ohmic Caldeira--Leggett model. For non-Ohmic environments, the reduced dynamics generally becomes non-Markovian owing to the emergence of memory kernels in the generalized Langevin equation. Consequently, the reduced density matrix, probability current, and the associated effective potential are expected to acquire memory-dependent corrections. Although the Bohmian construction can, in principle, be extended whenever a consistent probability current is available, the resulting trajectories would generally reflect the non-Markovian character of the environment. Exploring these effects constitutes an interesting direction for future investigation.

Overall, the present work provides a coherent framework in which dissipative dynamics, quantum-classical transition, and thermal effects are unified through the scaled Caldeira-Leggett structure. The emergence of classicality is governed not by stochastic forces acting on trajectories, but by the interplay between dissipation and the residual structure of the effective potential. This emergence of classicality is expected to be robust with respect to the choice of the initial state and is not restricted to Gaussian wave packets, although the latter allow for a fully analytical treatment within the CL formalism. It would be interesting to extend this approach to more complex environments and external potentials, and to explore other measures of nonclassicality within the same scaled formulation. Exploring possible extensions of the present framework to biological systems in particular sensory perception may constitute an interesting direction for future work. At this stage, however, we avoid from making any specific claims regarding such applications.

\vspace{1cm}

\vspace{0.5cm}
\noindent
{\bf Data availability}: This manuscript has no associated data.

\vspace{0.5cm}
\noindent
{\bf Acknowledgement}:
SVM gratefully acknowledges the support of the University of Qom. S. M.-A. acknowledges support of a grant from the Ministry of Science, Innovation and Universities (Spain) with Ref. PID2023-149406NB-I00. 
We thank R. Hamani and M. S. Kavyani for valuable discussions.



\appendix

\section{Derivation of the Newtonian-like equation}

In this appendix, we provide a detailed derivation of the Newtonian-like equation \eqref{eq: Newton}. Starting from the velocity field \eqref{eq: bohm-vel}, one has
\begin{eqnarray} \label{eq: ddotR_app}
m \ddot{x} &=& \frac{d}{dt} S_r \big|_{r=0, R=x}.
\end{eqnarray}
The total time derivative of $S_r$ reads
\begin{eqnarray} \label{eq: dSr_dt_app}
\frac{d}{dt} S_r &=& \left( \frac{\pa}{\pa t} + \dot{R} \frac{\pa}{\pa R} \right) S_r 
= \frac{\pa}{\pa r} \frac{\pa S}{\pa t} + \dot{R} S_{rR} \nonumber \\
&=& - \frac{\pa}{\pa r} \left( \frac{1}{m} S_r S_R + 2 \ga r S_r + \mathcal{V+Q} \right) + \frac{1}{m} S_r S_{rR} \nonumber \\
&=& -\frac{1}{m} S_{rr} S_R - 2 \ga S_r - 2 \ga r S_{rr} - \frac{\pa}{\pa r} ( \mathcal{V+Q} ),
\end{eqnarray}
where in the second line we have used the Hamilton-Jacobi equation \eqref{eq: HJ} and the definition of the velocity field, $\dot{R}=S_r/m$ according to \eqref{eq: bohm-vel}. 

The derivative of the potential term $\mathcal{V}$ is given by
\begin{eqnarray*}
\frac{\pa \mathcal{V}}{\pa r} &=& 
\frac{\pa V}{\pa x} \frac{\pa x}{\pa r} - \frac{\pa V}{\pa x'} \frac{\pa x'}{\pa r} 
= \frac{1}{2} \left( \frac{\pa V}{\pa x} + \frac{\pa V}{\pa x'} \right),
\end{eqnarray*}
and thus for the diagonal elements ($r=0$)
\begin{eqnarray*}
\frac{\pa \mathcal{V}}{\pa r} \bigg|_{r=0} &=& 
\frac{1}{2} \left( \frac{\pa V}{\pa x} + \frac{\pa V}{\pa x'} \right) \bigg|_{x'=x}
= \frac{\pa V}{\pa x},
\end{eqnarray*}
where in the first equality we have used the definition \eqref{eq: V-mat}.

It is worth noting that if the external potential admits a power series expansion, $V(x) = \sum_n c_n x^n$, then $\mathcal{V}(r,R)$ is an odd function of $r$:
\begin{eqnarray*}
\mathcal{V} &=& \sum_n c_n (x^n - x'^n) = \sum_n c_n \left[ (R + r/2)^n - (R - r/2)^n \right] \\
&=& \sum_n c_n \sum_{k=0}^{n} 
\begin{pmatrix} n \\ k \end{pmatrix} \frac{1 - (-1)^k}{2^k} r^k R^{n-k},
\end{eqnarray*}
implying that its derivative with respect to $r$, which corresponds to the force, is even in $r$.

For the effective potential, using Eqs. \eqref{eq: cm} and \eqref{eq: rel} in Eq. \eqref{eq: Q-mat} gives
\begin{eqnarray} \label{eq: Q-mat_app}
\mathcal{Q} = - \frac{\hb^2}{2m} \frac{1}{A(x,x',t)} \left( \frac{\pa^2}{\pa x^2} - \frac{\pa^2}{\pa x'^2} \right) A(x,x',t),
\end{eqnarray}
where $A$ is the amplitude of the density matrix. Noting that $A(x,x',t)$ is even in $x \leftrightarrow x'$, it follows that $\mathcal{Q}$ is odd under this interchange. Hence, the derivative with respect to $r$ at $r=0$ becomes
\begin{eqnarray} \label{eq: eff-force}
\frac{\pa \mathcal{Q}}{\pa r} \bigg|_{r=0} &=& 
\frac{1}{2} \left( \frac{\pa}{\pa x} - \frac{\pa}{\pa x'} \right) \mathcal{Q}(x,x',t) \bigg|_{x'=x} 
= \frac{\pa \mathcal{Q}}{\pa x} \bigg|_{x'=x}.
\end{eqnarray}

Finally, from the oddness of the phase function $S(r,R,t)$ in $r$ (Eq. \eqref{eq: S-odd}), one has $S_{rr}\big|_{r=0} = 0$. Collecting all these results into Eqs. \eqref{eq: ddotR_app} and \eqref{eq: dSr_dt_app}, we recover the Newtonian-like equation \eqref{eq: Newton}.

\end{document}